\documentclass[conference,a4paper]{IEEEtran}

\usepackage{bm}
\usepackage[cmex10]{amsmath}
\usepackage{amssymb}
\usepackage{mathrsfs} 
\DeclareMathAlphabet{\mathscrbf}{OMS}{mdugm}{b}{n}
\usepackage{nicefrac}
\usepackage[space,compress,sort]{cite}
\usepackage{mathtools}
\usepackage{theorem}
\usepackage{mathrsfs}
\usepackage{centernot}
\usepackage{hyperref}
\usepackage[nodisplayskipstretch]{setspace}
\usepackage{multirow} 

\newcommand{\ie}{i.e.,\ }
\newcommand{\eg}{e.g.,\ }
\newcommand{\defeq}{\ensuremath{\triangleq}}
\renewcommand{\vec}[1]{\ensuremath{\boldsymbol{#1}}}
\newcommand{\mat}[1]{\ensuremath{\boldsymbol{#1}}}
\newcommand{\q}{\ensuremath{q}}
\newcommand{\F}{\ensuremath{\mathbb{F}}}
\newcommand{\Fq}{\ensuremath{\F_{\q}}}
\newcommand{\N}{\ensuremath{\mathbb{N}}}
\newcommand{\PERf}[1]{\ensuremath{\mathscr{P}_{#1}}}
\newcommand{\PER}[2]{\ensuremath{\PERf{#1}}\left[#2\right]}
\newcommand{\REENCf}[1]{\ensuremath{\mathscr{R}_{#1}}}
\newcommand{\REENC}[2]{\ensuremath{\REENCf{#1}\left[#2\right]}}
\newcommand{\DFT}[1]{\ensuremath{\mathscr{F}\left[#1\right]}}
\newcommand{\IDFT}[1]{\ensuremath{\mathscr{F}^{-1}\left[#1\right]}}
\newcommand{\RS}{\ensuremath{\mathcal{RS}}}
\newcommand{\RSpar}{\ensuremath{\RS\left(\Fq; n, k, d\right)}}
\newcommand{\Hw}[1]{\ensuremath{\mathrm{wt_\mathrm{H}}\left[#1\right]}}
\newcommand{\Hd}[2]{\ensuremath{\mathrm{d_\mathrm{H}}\left[#1, #2\right]}}

{\theoremstyle{plain}
\newtheorem{definition}{Definition}}

{\theoremstyle{plain}
\newtheorem{theorem}{Theorem}}

{\theoremstyle{plain}
\newtheorem{corollary}{Corollary}}

{\theoremstyle{plain}
\newtheorem{problem}{Problem}}

\IEEEoverridecommandlockouts

\begin{document}

\sloppy

\title{Re-Encoding Techniques for Interpolation-Based Decoding of Reed--Solomon Codes}

\author{
  \IEEEauthorblockN{Christian Senger}\thanks{This work has been supported by DFG,
Germany, under grant BO~867/22-1.}
  \IEEEauthorblockA{Institute of Communications Engineering, Ulm University, Ulm, Germany\\
    Email: christian.senger@uni-ulm.de}
}

\maketitle

\begin{abstract}
We consider interpolation-based decoding of Reed--Solomon codes using the Guruswami--Sudan algorithm (GSA) and investigate the effects of two modification techniques for received vectors, \ie the re-encoding map and the newly introduced periodicity projection. After an analysis of the latter, we track the benefits (that is low Hamming weight and regular structure) of modified received vectors through the interpolation step of the GSA and show how the involved homogeneous linear system of equations can be compressed. We show that this compression as well as the recovery of the interpolated bivariate polynomial is particularly simple when the periodicity projection was applied.
\end{abstract}

\section{Introduction}\label{sec:intro}

The discovery of the \emph{Guruswami--Sudan list decoding algorithm (GSA)} \cite{guruswami_sudan:1999} was one of the major breakthroughs in algebraic coding theory. Building up upon the original \emph{Sudan list decoding algorithm} \cite{sudan:1997}, which is restricted to the practically less interesting case of low code rates, it manages to decode Reed--Solomon codes (and, more generally, Algebraic Geometry codes) of arbitrary rate beyond half their minimum distance with polynomial time complexity in the code length. It was pointed out in \cite{guruswami_sudan:1999} that the two algorithms are closely related to each other and that the older algorithm is a special case of the newer one. It can be attributed to the work of Gemmell and Sudan \cite{gemmell_sudan:1992} that the classical bounded minimum distance decoder of Welch and Berlekamp \cite{welch_berlekamp:1986} is a special case of both. The latter fact is nicely elaborated in \cite[Sections~5.2 and 12.2]{justesen_hoholdt:2004}.

Guruswami and Sudan did not aim to reduce the degree of the complexity polynomial and, as we will see in Section~\ref{sec:dec}, their original algorithm is in $\mathcal{O}\left[s^3\ell^3 n^3\right]$, where $n$ is the code length and $s, \ell$ are two interdependent parameters to be explained later. Since 1999, several authors proposed improved versions of the algorithm, whose time complexities are quadratic in the code length instead of cubic. Most improvements are based on the fact that the computationally most expensive task of the algorithm is bivariate polynomial interpolation, which can be realized as a linear system of equations. The complexity of solving general linear systems using Gaussian elimination is cubic in their size, but it has been observed that the linear system which appears in the GSA is a rather structured one.

Among the fastest realizations of the interpolation is the solution of to Augot and Zeh \cite{augot_zeh:2008}, see also \cite{zeh_gentner_bossert:2009}. It is based on an adaptation of Feng and Tzeng's \emph{Fundamental Iterative Algorithm (FIA)} \cite{feng_tzeng:1991} for block Hankel matrices, which is in turn due to Roth and Ruckenstein \cite{roth_ruckenstein:2000}. Its time complexity is in $\mathcal{O}\left[l s^4 n^2\right]$. Other solutions have been presented by Alekhnovich \cite{alekhnovich:2002} (Diophantine equations), Olshevsky and Shokrollahi \cite{olshevsky_shokrollahi:2003} (matrix displacement/Schur complement), and Trifonov \cite{trifonov:2010} (Gr\"obner bases). More references can be found in \cite[Chapter~9]{roth:2006}.

Another technique to reduce the complexity of the GSA is to reduce the size of the enclosed interpolation problem. This \emph{re-encoding} approach was followed by Gross et~al.\ \cite{gross_kschischang_koetter_gulak:2003}, Ma \cite{ma:2007}, and K\"otter et~al.\ \cite{koetter_ma_vardy:2011}. In their papers, bivariate interpolation is done using polynomial-based algorithms like the K\"otter algorithm \cite{koetter:1996}.

Our contribution is the description and analysis of a particularly simple case of re-encoding --- the \emph{periodicity projection} --- which allows to compress the size of the interpolation problem using especially sparse data structures while at the same time it maintains its regular structure. We conjecture that this is useful for the complete GSA, but in this paper we focus only on the involved interpolation. Along the way, we investigate the effect of general re-encoding on the linear system and make it applicable for matrix-based interpolation as in \cite{augot_zeh:2008, zeh_gentner_bossert:2009}.

The rest of the paper is organized as follows. In Section~\ref{sec:basics} we define some basic notions and recall the re-encoding map. Section~\ref{sec:PER} is devoted to periodic vectors and the periodicity projection. Besides their fundamental properties we also explain their relation to the re-encoding map. In Section~\ref{sec:dec} we shortly recapitulate the GSA, before we investigate the GSA when it is applied to sparse and structured received vectors as created by the re-encoding map and the periodicity projection in Section~\ref{sec:compress} and show how it can be sped up. Section~\ref{sec:conc} concludes the paper with some closing comments.

\section{Reed--Solomon Codes and Re-Encoding}\label{sec:basics}

\begin{definition}\label{def:DFT}
  Let $\q$ be a prime power, $n\defeq\q-1$, and let $\Fq$ be a finite field with $\q$ elements. Let further $\alpha\in\Fq$ be a primitive element. The \emph{Discrete Fourier Transform (DFT)} of $\vec{v}=(v_0, \ldots, v_{n-1})\in\Fq^n$ is $\DFT{\vec{v}}=\vec{V}=(V_0, \ldots, V_{n-1})\in\Fq^n$, where
  \begin{equation*}
    V_j\defeq \sum_{i=0}^{n-1}v_i \alpha^{ji},
  \end{equation*}
  and $\vec{V}$ is denoted as a \emph{frequency-domain vector}. The \emph{Inverse Discrete Fourier Transform (IDFT)} of $\vec{V}$ is $\IDFT{\vec{V}}=\vec{v}$, and it holds
  \begin{equation}\label{eqn:idft}
    v_j\defeq n^{-1} \sum_{i=0}^{n-1}V_i \alpha^{-ji}.
  \end{equation}
  The vector $\vec{v}$ is referred to as a \emph{time-domain vector}.
\end{definition}

Note that the DFT can be interpreted as the vector-matrix multiplication $\DFT{\vec{v}}=\vec{v}\mathscrbf{F}$ with an $n\times n$ \emph{Vandermonde matrix} $\mathscrbf{F}\defeq\left(\alpha^{-ji}\right)_{j, i}$. Using this interpretation, the IDFT becomes $\IDFT{\vec{V}}=\vec{V}\mathscrbf{F}^{-1}$.

\begin{definition}\label{def:RS}
  For parameters $n, k\in\N$ with $k\leq n\defeq q-1$, a primitive \emph{Reed--Solomon (RS) code} over $\Fq$ can be defined as\vspace{-0.3cm}
  \begin{multline*}
    \RSpar\defeq\\
    \left\{%
      \IDFT{\vec{C}}:\vec{C}=(\underbrace{C_0, \ldots, C_{k-1}}_{k\;\text{times}}, \underbrace{0, \ldots, 0}_{n-k\,\text{times}}), C_j\in\Fq
    \right\}.
  \end{multline*}
  \emph{Length} and \emph{dimension} of the code are given by $n$ and $k$, respectively. $\RS$ is linear, \ie for $\beta_1, \beta_2\in\Fq$ holds the implication $\vec{c}_1\vec{c}_2\in\RS\Longrightarrow\beta_1\vec{c}_1+\beta_2\vec{c}_2\in\RS$.
\end{definition}

It is a well-known fact that for the \emph{minimum distance} $d$ of RS codes holds equality in the Singleton Bound, \ie $d=n-k+1$. Since the minimum distance of a code is defined as the minimal number of positions in which any two codewords $\vec{c}_1$ and $\vec{c}_2$, $\vec{c}_1\neq\vec{c}_2$, differ, it immediately follows that any codeword is uniquely determined by any $k$ of its positions. This is commonly denoted as the \emph{Maximum Distance Separable (MDS)} property.

Transmission of a codeword $\vec{c}\in\RS$ over a channel results in the reception of a received vector $\vec{r}=\vec{c}+\vec{e}$, which is distorted by an error vector $\vec{e}$. Inspired by the MDS property, we can define the following simple mapping. It allows to map any vector $\vec{r}$ received from the transmission channel to a modified received vector with beneficial properties for the decoding.

\begin{definition}\label{def:REENC}
   For an RS code $\RSpar$ let $\mathcal{J}=\left\{j_0, \ldots, j_{\sigma-1}\right\}$ be a set of $\sigma$ positions, $0\leq j_t\leq n-1$. Then the \emph{re-encoding} map with regard to $\mathcal{J}$ is
  \begin{equation*}
    \REENCf{\mathcal{J}} ~:~ \left\{\begin{array}{rcl} \Fq^n &\to &\Fq^n\\
    \vec{v} &\mapsto &\vec{v}+\widetilde{\vec{c}}\end{array}\right.
  \end{equation*}
  where $\widetilde{\vec{c}}=\left(\widetilde{c}_0, \ldots, \widetilde{c}_{n-1}\right)\in\RS$, such that for all $j\in\mathcal{J}$ holds $v_j=\widetilde{c}_j$.
\end{definition}

If $\sigma=k$ (the only practically relevant case), then $\REENCf{\mathcal{J}}$ is a projection, \ie it is idempotent. This can be seen by the fact that all positions $j\in\mathcal{J}$ in $\REENC{\mathcal{J}}{\vec{v}}=\vec{v}+\widetilde{\vec{c}}_1$ are zero by definition. But then $\REENC{\mathcal{J}}{\REENC{\mathcal{J}}{\vec{v}}}=\vec{v}+\widetilde{\vec{c}}_1+\widetilde{\vec{c}}_2$, where $\widetilde{\vec{c}}_2\in\RS$ is zero at the $k$ positions in $\mathcal{J}$, which is only possible if it is the all-zero codeword.

The effect of re-encoding when applied to a received vector $\vec{r}$ is straightforward to see. It maps $\vec{r}=\vec{c}+\vec{e}$ to
\begin{equation*}
  \REENC{\mathcal{J}}{\vec{r}}=\underbrace{\vec{c}+\widetilde{\vec{c}}}_{\defeq\vec{c}^\prime\in\RS}+\vec{e},
\end{equation*}
\ie to another received vector with at most $n-\sigma$ non-zero positions. We will see in Section~\ref{sec:dec} why this is useful. It should be clear that a codeword $\widetilde{\vec{c}}$ can be efficiently calculated using an erasures-only decoder as long as $\sigma\leq n-k$. It should also be clear that the transmitted codeword $\vec{c}$ can be recovered after successful decoding of the modified received vector $\REENC{\mathcal{J}}{\vec{r}}$.

Re-encoding is a well-known concept in algebraic coding. It has been used implicitly by Welch and Berlekamp in \cite{welch_berlekamp:1986} in order to decrease the complexity of their interpolation-based bounded minimum distance decoder and has been revitalized in the context of interpolation-based list decoding \cite{gross_kschischang_koetter_gulak:2003, ma:2007, koetter_ma_vardy:2011}.

\section{The Periodicity Projection}\label{sec:PER}

This section is devoted to a mapping technique for received vectors first proposed in \cite{senger:2012}. It is based on a certain property of the DFT, which we recall as Theorem~\ref{thm:zeropos}. We show at the end of the section that this technique is a special case of re-encoding as in Definition~\ref{def:REENC}.

\begin{definition}
  Let $n\defeq \q-1$ for a prime power $\q$ and let $p\in\N\setminus\{0\}$ such that $p\mid n$. A vector $\vec{V}\in\Fq^n$ of the form
  \begin{equation*}
    \vec{V}=(\underbrace{\vec{T}, \ldots, \vec{T}}_{\nicefrac{n}{p}\;\text{times}}),
  \end{equation*}
  where $\vec{T}$ is a \emph{template vector} of the form
  \begin{equation*}
    \vec{T}=(T_0, \ldots, T_{p-1}),
  \end{equation*}
  is denoted as a \emph{$p$-periodic vector}.
\end{definition}

In this paper, all $p$-periodic vectors are frequency domain vectors without particularly mentioning it. The following theorem relates $p$-periodicity in frequency domain to sparsity in time domain. It can be seen as a less prominent property of the DFT next to widely known properties as, \eg the \emph{convolution property} \cite[Theorem~6.1.3]{blahut:2003} or the \emph{polynomial root property} \cite[Theorem~6.1.5]{blahut:2003}.

\begin{theorem}\label{thm:zeropos}
  A vector $\vec{V}=(V_0, \ldots, V_{n-1})$ is $p$-periodic with template vector $\vec{T}=(T_0, \ldots, T_{p-1})$ if and only if its time-domain counterpart is $\vec{v}=(v_0, \ldots, v_{n-1})=\IDFT{\vec{V}}\in\Fq^n$, where
  \begin{equation*}
    v_j=\left\{\begin{array}[]{ll}
                 \displaystyle p\sum_{s=0}^{p-1} T_s \alpha^{-sj} & \text{if}\; \frac{n}{p}\mid j\\
                 0 & \text{if}\; \frac{n}{p}\centernot\mid j
               \end{array}\right..
  \end{equation*}
\end{theorem}
\begin{IEEEproof}
  By Definition~\eqref{eqn:idft} $v_j= n^{-1}\sum_{i=0}^{n-1}V_i \alpha^{-ij}$, which can be written in terms of the template vector as
  \begin{equation}\label{eqn:vjintermsofT}
    v_j=n^{-1}\sum_{t=0}^{\nicefrac{n}{p}-1}\sum_{s=0}^{p-1}T_s\left(\alpha^{-j}\right)^{t p+s}.
  \end{equation}
  Let us write $j=\nicefrac{\mu n}{p}+\nu$ with $\mu,\nu\in\N$. Then the summand can be written as
  \begin{align}
    T_s\left(\alpha^{-j}\right)^{t p+s}%
    &= T_s\left(\alpha^{-\nicefrac{\mu n}{p}-\nu}\right)^{t p+s}\nonumber\\
    &= T_s\alpha^{-\nicefrac{\mu n t p}{p}-\nicefrac{\mu n s}{p} -\nu t p-\nu s}\nonumber\\
    &=\alpha^{-\nu t p} T_s\alpha^{-\mu t(q-1)-s(\nicefrac{\mu n}{p}+\nu)}\nonumber\\
    &\stackrel{(*)}{=}\alpha^{-\nu t p} T_s\alpha^{-s(\nicefrac{\mu n}{p}+\nu)},\label{eqn:summandgeneral}
  \end{align}
  where $(*)$ follows from the fact that for any $\beta\in\Fq$ holds $\beta^{\q-1}=1$ (Lagrange's Theorem). In case $\nicefrac{n}{p}\mid j$, we have $\nu=0$ and $j=\nicefrac{\mu n}{p}$. Hence, using \eqref{eqn:summandgeneral}, \eqref{eqn:vjintermsofT} becomes
  \begin{equation*}
    v_j=n^{-1}\sum_{t=0}^{\nicefrac{n}{p}-1}\sum_{s=0}^{p-1}T_s\alpha^{-sj}=p\sum_{s=0}^{p-1}T_s\alpha^{-sj},
  \end{equation*}
  which proves the first part of the statement. If $\nicefrac{n}{p}\centernot\mid j$, then $\nu\in\N\setminus\{0\}$ and, again using \eqref{eqn:summandgeneral}, \eqref{eqn:vjintermsofT} becomes
  \begin{equation*}
    v_j=n^{-1}\sum_{t=0}^{\nicefrac{n}{p}-1}\alpha^{-\nu t p}\sum_{s=0}^{p-1}T_s\alpha^{-s(\mu a+\nu)}\,
  \end{equation*}
  where we can exchange inner and outer summation in order to obtain
  \begin{equation*}
    v_j=n^{-1}\sum_{s=0}^{p-1}T_s%
      \alpha^{-s(\nicefrac{\mu n}{p}+\nu)}\sum_{t=0}^{\nicefrac{n}{p}-1}\alpha^{-\nu t b}.
  \end{equation*}
  But
  \begin{equation}\label{eqn:geometricseries}
    \sum_{t=0}^{\nicefrac{n}{p}-1}\left(\alpha^{-\nu p}\right)^t=\frac{\left(1-\alpha^{-n\nu}\right)}{\left(1-\alpha^{-p\nu}\right)}=\frac{\left(1-\alpha^{-(\q-1)\nu}\right)}{\left(1-\alpha^{-p\nu}\right)}=0,
  \end{equation}
  since the sum is a geometric series, proving that $v_j=0$ whenever $\nicefrac{n}{p}\centernot\mid j$.
\end{IEEEproof}

It is interesting to note that periodicity is maintained by cyclic convolution with arbitrary vectors. This simple corollary can be easily seen using the convolution property of the DFT.

\begin{corollary}\label{cor:prodperiodic}
  Let $\vec{V}=(V_0, \ldots, V_{n-1})\in\Fq^n$ be $p$-periodic and $\vec{W}=(W_0, \ldots, W_{n-1})\in\Fq^n$ be arbitrary. Then the cyclic convolution $\vec{U}=(U_0, \ldots, U_{n-1})$ of these vectors, where
  \begin{equation*}
    U_j\defeq \sum_{i=0}^{j} V_{j-i}W_i=\sum_{i=0}^{j} W_{j-i}V_i,
  \end{equation*}
  is $p$-periodic.
\end{corollary}

\begin{definition}\label{def:PER}
  Let $p\in\N\setminus\{0\}$ such that $p\mid n$. Then the \emph{periodicity projection} with regard to $p$ is defined as the map
  \begin{equation*}
    \PERf{p} ~:~ \left\{\begin{array}{rcl} \Fq^n &\to &\Fq^n\\
    \vec{v} &\mapsto & \mathscr{F}^{-1}[(\underbrace{\vec{T}, \ldots, \vec{T}}_{\nicefrac{n}{p}\;\text{times}})]\end{array}\right.,
  \end{equation*}
  where $\vec{T}=(V_{n-p}, \ldots, V_{n-1})$ and $\DFT{\vec{v}}=(V_0, \ldots, V_{n-1})$.
\end{definition}

The map $\PERf{p}$ is indeed a projection, \ie it is idempotent. This follows from the fact that its values depend only on the rightmost $p$ positions of the frequency-domain counterpart of the input vector $\vec{v}$ and these positions are not affected by $\PERf{p}$. Note that the periodicity projection can be interpreted as a linear operator $\mathscrbf{P}=\mathscrbf{F}^{-1}\mat{P}\mathscrbf{F}$ with a sparse matrix
\begin{equation*}\setstretch{1.5}
  \mat{P}\defeq\left(\begin{array}{c|c}
    \multirow{3}{*}{\mbox{\huge{$\mat{0}$}\tiny{$_{n\times (n-p)}$}}}
    & \mbox{\footnotesize{$\mat{1}$}\tiny{$_{p\times p}$}}\\\cline{2-2}
    & \mbox{\tiny{$\vdots$}}\\\cline{2-2}
    & \mbox{\footnotesize{$\mat{1}$}\tiny{$_{p\times p}$}}\\
    \end{array}\right)\in\Fq^{n\times n}.
\end{equation*}
Here, $\mat{0}$ denotes the all-zero matrix and $\mat{1}$ the identity matrix.

\begin{theorem}\label{thm:RSperiodic}
  Let $\RSpar$ be an RS code. Let further $p\in\N\setminus\{0\}$ such that $p\mid n$, and $p\geq d-1$. If $\vec{c}\in\RS$ is a codeword, $\vec{e}\in\Fq^n$ is an error vector of Hamming weight $\Hw{\vec{e}}=\varepsilon$, and $\vec{r}=\vec{c}+\vec{e}$ is the  received vector, then
  \begin{align*}
    \PER{p}{\vec{r}}&=\vec{c}^\prime+\vec{e},\\
    \shortintertext{where}
    \vec{c}^\prime&\defeq \vec{c}+\PER{p}{\vec{r}}-\vec{r}
  \end{align*}
  and $\vec{c}^\prime\in\RS$ with $\Hw{\vec{c}^\prime}\le p+\varepsilon$.
\end{theorem}
\begin{IEEEproof}
  Let $\vec{R}=(R_0, \ldots, R_{n-1})\DFT{\vec{r}}$ be the frequency-domain counterpart of $\vec{r}$. By Definition~\ref{def:PER}, 
  \begin{equation}\label{eqn:conc}
    \PER{p}{\vec{r}}=\mathscr{F}^{-1}[(\underbrace{\vec{T}, \ldots, \vec{T}}_{\nicefrac{n}{p}\;\text{times}})],
  \end{equation}
  where the template vector $\vec{T}=(R_{n-p}, \ldots, R_{n-1})$ consists of the $p$ rightmost components of $\vec{R}$. If we group the components of $\vec{r}$ into $\nicefrac{n}{p}$ blocks of length $p$, \ie $ \vec{R}=(\vec{R}_0, \ldots, \vec{R}_{\nicefrac{n}{p}-1})$, then $\vec{T}=\vec{R}_{\nicefrac{n}{p}-1}$ and consequently
  \begin{equation*}
    \PER{p}{\vec{r}}-\vec{r} = \mathscr{F}^{-1}[(\vec{T}-\vec{R}_0, \ldots, \vec{T}-\vec{R}_{\nicefrac{n}{p}-2}, \vec{Z})],
  \end{equation*}
  where $\vec{Z}$ is the all-zero vector of length $p$. Since we assumed $p\geq d-1$, it follows from Definition~\ref{def:RS} that $\widetilde{\vec{c}}= \PER{p}{\vec{r}}-\vec{r}\in\RS$. The first part of the claim follows from substituting $\vec{r}=\vec{c}+\vec{e}$ and the linearity of $\RS$. As for the second claim, it follows from Theorem~\ref{thm:zeropos} that $\Hw{\PER{p}{\vec{r}}}\leq p$. But $\PER{p}{\vec{r}}=\vec{c}^\prime+\vec{e}$, $\Hw{\vec{e}}=\varepsilon$. Thus, $\vec{c}^\prime$ differs from $\PER{p}{\vec{r}}$ in at most $\varepsilon$ positions and the bound $\Hw{\vec{c}^\prime}\le p+\varepsilon$ follows.
\end{IEEEproof}

We emphasize that for $\vec{r}=\vec{c}+\vec{e}$ and $\PER{p}{\vec{r}}=\vec{c}^\prime+\vec{e}$ as in the theorem generally holds $\PER{p}{\vec{c}}\neq\vec{c}^\prime$ and $\PER{p}{\vec{e}}\neq\vec{e}$. From Theorem~\ref{thm:zeropos} and the proof of Theorem~\ref{thm:RSperiodic} we can immediately extract the following simple corollary.

\begin{corollary}\label{cor:specialcase}
  The periodicity projection is a special case of the re-encoding map with $\sigma=n-p$ and $\mathcal{J}=\left\{j:\nicefrac{n}{p}\centernot\mid j\right\}$.
\end{corollary}

\section{The Guruswami--Sudan Algorithm}\label{sec:dec}

The GSA \cite{guruswami_sudan:1999} can be divided into two steps, the interpolation step (Problem~\ref{prob:interpolation}) and the factorization step (Problem~\ref{prob:factorization}). Our focus here is on the interpolation step, which is computationally more involved. Let in the following $\vec{c}\in\RSpar$ be a codeword, $\vec{e}\in\Fq^n$ be an error vector of Hamming weight $\Hw{\vec{e}}=\varepsilon$, and $\vec{r}=\vec{c}+\vec{e}$ be a received vector from the transmission channel. Furthermore, let $s,\ell\in\N\setminus\{0\}$ be two parameters of the GSA with $s<\ell$.

\begin{problem}[Interpolation Step]\label{prob:interpolation}
  Find a non-zero bivariate polynomial $Q(x, y)=Q_0(x)+Q_1(x)y+\cdots +Q_\ell(x)y^\ell$ over $\Fq$ such that
  \begin{equation}\label{eqn:degbound}
    \deg\left[Q_\nu(x)\right]\leq s(n-\varepsilon)-1-\nu(k-1)\defeq d_\nu
  \end{equation}
  and\vspace{-0.3cm}
  \begin{multline}\label{eqn:hasse}
    \forall j=0, \ldots, n-1\;\text{and}\;a,b\in\N, a+b<s:\\
    \sum_{\nu=b}^{\ell}\sum_{\mu=a}^{d_\nu}
    \binom{\mu}{a}\binom{\nu}{b}Q_{\mu, \nu}x^{\mu-a}y^{\nu-b}%
    \bigg|_{(x, y)=\left(\alpha^{-j}, r_j\right)}=0,
  \end{multline}
  where $Q_\nu(x)=\sum_{\mu=0}^{d_\nu} Q_{\mu, \nu}x^\mu$.
\end{problem}

The nested sum in \eqref{eqn:hasse} is called the $(a, b)$-th mixed partial Hasse derivative \cite{hasse:1936} of $Q(x, y)$. The condition that all $(a, b)$-th Hasse derivatives with $a+b<s$ evaluate to zero for all tuples $(\alpha^{-j}, r_j)$, $j=0, \ldots, n-1$, means by that these tuples are zeros of multiplicity $s$ of $Q(x, y)$. For that reason, we refer to the parameter $s$ as the \emph{multiplicity} of the GSA. It can be shown that the homogeneous linear system associated with Problem~\ref{prob:interpolation} has a non-zero solution (\ie it has more equations than unknowns) as long as
\begin{equation*}
  \varepsilon<\varepsilon_0\defeq\frac{n(2\ell-s+1)}{2(\ell+1)}-\frac{\ell(k-1)}{2s}.
\end{equation*}

A quick analysis shows that the system has $\nicefrac{n s(s+1)}{2}$ equations and $\sum_{\nu=0}^{\ell}\left(d_\nu+1\right)$ unknowns. Both numbers are exceedingly large even for short RS codes and intermediate parameters $s$ and $\ell$. As a result, the time complexity of solving the system with Gaussian elimination is in $\mathcal{O}\left[s^3\ell^3 n^3\right]$.

After a solution of Problem~\ref{prob:interpolation} is found, the following problem must be solved.

\begin{problem}[Factorization Step]\label{prob:factorization}
  Given a solution $Q(x, y)$ of Problem~\ref{prob:interpolation}, find all factors $y-F(x)$ with $\deg\left[F(x)\right]<k$.
\end{problem}

If we associate the at most $\ell$ resulting polynomials $F_\kappa(x)$ with padded vectors
\begin{equation*}
  \vec{F}_\kappa=(F_{\kappa, 0}, \ldots, F_{\kappa, k-1}, \underbrace{0, \ldots, 0}_{n-k\;\text{times}}),
\end{equation*}
then Definition~\ref{def:RS} tells that $\vec{f}_\kappa\defeq\IDFT{\vec{F}_\kappa}\in\RS$. Since the parameters were chosen such that the GSA can correct at most $\varepsilon_0$ errors, the \emph{result list} $\mathcal{L}$ of the GSA contains all $\vec{f}_\kappa$ with $\Hd{\vec{f}_\kappa}{\vec{r}}\leq\varepsilon_0$. It is proven in \cite{guruswami_sudan:1999} that under all these assumptions $\vec{c}\in\mathcal{L}$. Since $\mid\mathcal{L}\mid\leq\ell$, we refer to $\ell$ as the \emph{list size} of the GSA.

Problem~\ref{prob:factorization} can be solved with time complexity in $\mathcal{O}\left[l \log\log[l]n^2\right]$ using a technique from \cite{roth_ruckenstein:2000}, but this is not within the scope of this manuscript.

It follows from  the exposition of the Welch--Berlekamp algorithm in \cite{gemmell_sudan:1992} and the interpretation of Justesen and H\o holdt in \cite[Sections~5.2 and 12.2]{justesen_hoholdt:2004} that the GSA simplifies to the Sudan algorithm if we restrict the multiplicity to $s=1$ and that it further simplifies to the Welch--Berlekamp algorithm if we additionally restrict the list size to $\ell=1$.

\section{Compressing the Interpolation Step}\label{sec:compress}

We will now investigate the GSA when it is applied to a modified received vector $\REENC{\mathcal{J}}{\vec{r}}$ regarding $\mathcal{J}$. Doing so, we examine the Hasse derivative in \eqref{eqn:hasse} for $j\in\mathcal{J}$. Slightly misusing mathematical notation, it becomes
\vspace{-0.3cm}
\begin{multline*}
  \forall j\in\mathcal{J}\;\text{and}\;a,b\in\N, a+b<s:\\
  \sum_{\nu=b}^{\ell}\sum_{\mu=a}^{d_\nu}
  \binom{\mu}{a}\binom{\nu}{b}Q_{\mu, \nu}x^{\mu-a}0^{\nu-b}%
  \bigg|_{x=\alpha^{-j}}=0,
\end{multline*}
meaning that the value of the inner sum is non-zero if and only if $\nu=b$. But this allows to write
\vspace{-0.3cm}
\begin{multline*}
  \forall j\in\mathcal{J}\;\text{and}\;a,b\in\N, a<s-b:\\
  \sum_{\mu=a}^{d_b}
  \binom{\mu}{a}Q_{\mu, b}x^{\mu-a}%
  \bigg|_{x=\alpha^{-j}}=0
\end{multline*}
which means that $\alpha^{-j}$ is a root of multiplicity $s-b-1$ of all $Q_b(x)$, $0\leq b<s$. Let us define the polynomial
\begin{equation}\label{eqn:Vx}
  V(x)\defeq\prod_{j\in\mathcal{J}} (x-\alpha^{-j}).
\end{equation}

With that we have proven the first part of our first main theorem:

\begin{theorem}
  Let $Q(x, y)=Q_0(x)+Q_1(x)y+\cdots +Q_\ell(x)y^\ell$ be a solution of Problem~\ref{prob:interpolation} for a modified received vector $\REENC{\mathcal{J}}{\vec{r}}$ with regard to $\mathcal{J}$. Then the $Q_b(x)$, $0\leq b<s$, can be written as
  \begin{equation*}
    Q_b(x)=W_b(x)V_b(x),
  \end{equation*}
  where $V_b(x)\defeq V(x)^{s-b}$ and $V(x)$ is known and given by \eqref{eqn:Vx} and
  \begin{equation*}
    \deg\left[W_b(x)\right]\leq d_b-\sigma(s-b).
  \end{equation*}
\end{theorem}
\begin{IEEEproof}
  The second part follows from the simple observation that $\deg\left[V(x)^{s-b}\right]=\sigma(s-b)$ and \eqref{eqn:degbound}.
\end{IEEEproof}

We will now interpret this result in the setting of the homogeneous linear system associated with \eqref{eqn:hasse}. The coefficients of $Q_b(x)=\sum_{\mu=0}^{d_b} Q_{\mu, b}x^\mu$, $b<s$, are a subset of the system's solution variables and we know from the theorem that they can be written as a linear combination with known factors (the coefficients of $V_b(x)=\sum_{\mu=0}^{\sigma(s-b)} V_{\mu, b}x^\mu$) and unknown terms (the coefficients of $W_b(x)=\sum_{\mu=0}^{d_b-\sigma(s-b)} W_{\mu, b}x^\mu$).

If it is known in advance that some of the solution variables of a system are linearly dependent, a simple and well-known trick can be applied. Due to space restrictions, we state this trick only by means of a small example. Assume that we have a linear system over an arbitrary field
\begin{equation*}
  \begin{pmatrix}
    a & b & c\\
    d & e & f\\
    g & h & i\\
  \end{pmatrix}%
  \cdot%
  \begin{pmatrix}
    x\\
    y\\
    z\\
  \end{pmatrix}%
  =%
  \begin{pmatrix}
    j\\
    k\\
    l\\
  \end{pmatrix}
\end{equation*}
and know that a solution $(x, y, z)^T$ exists and $z=\alpha x+\beta y$. Then, $(x, y, z)^T$ can as well be recovered by first solving
\begin{equation*}
  \begin{pmatrix}
    a+c\alpha & b+c\beta\\
    d+f\alpha & e+f\beta\\
    g+i\alpha & h+i\beta\\
  \end{pmatrix}%
  \cdot%
  \begin{pmatrix}
    x\\
    y\\
  \end{pmatrix}%
  =%
  \begin{pmatrix}
    j\\
    k\\
    l\\
  \end{pmatrix}
\end{equation*}
for $(x, y)^T$ and then setting $z=\alpha x+\beta y$. Obviously, at least one column of the compressed coefficient matrix is linearly dependent on the others.

Repeated and nested application of the trick basically allows to solve a linear system for the coefficients of the polynomials $W_0(x), \ldots, W_{s-1}(x), Q_s(x), \ldots, Q_\ell(x)$ instead of $Q_0(x), \ldots, Q_\ell(x)$, meaning that the number of unknowns --- and, in doing so, the system size --- is diminished by $\sum_{b=0}^{s-1}\sigma(s-b)$. This compression effect is maximized for larger $\sigma$ hence in a practical system one should always choose $\sigma=n-k$. In that case, the number of linearly independent equations in the system is diminished from $\nicefrac{n s(s+1)}{2}$ to $\nicefrac{(n-k)s(s+1)}{2}$. However, it is not a priori clear for general re-encoding which of the equations can be discarded due to linear dependence.

Now consider the GSA when applied to $\PER{p}{\vec{r}}$. Recall that this is equivalent to general re-encoding with $\sigma=n-p$ and $\mathcal{J}=\left\{j:\nicefrac{n}{p}\centernot\mid j\right\}$. In that case, it follows from Lagrange's Theorem that for each root $\alpha^{-j}$ of $V(x)$, its inverse $\alpha^j$ is as well a root of $V(x)$. Polynomials with this property are \emph{palindromic} or \emph{self-reciprocal}, \ie the sequence of their coefficients is a palindrome. This allows to prove our second main theorem.

\begin{theorem}
  Let $Q(x, y)=Q_0(x)+Q_1(x)y+\cdots +Q_\ell(x)y^\ell$ be a solution of Problem~\ref{prob:interpolation} for a modified received vector $\PER{p}{\vec{r}}$ with regard to $p$. Then $V_b(x)$, $0\leq b<s$, is palindromic with constant term $1$ and it is sparse, with non-zero coefficients displaced by at least $p$.
\end{theorem}

This structure of the $V_b(x)$, $0\leq b<s$, renders nested and repeated application of the trick as well as reconstruction of the complete polynomial $Q(x, y)$ from the solution of the compressed system nearly trivial, because the coefficients $Q_{\mu, b}$ are either sparse linear combinations of the $W_{\mu, b}$, simple multiples of a $W_{\mu, b}$, or even constant zero. Besides that, the palindromic structure yields that in the compressed coefficient matrix of the linear system, all rows $i$ with $\left(\nicefrac{n}{p}\centernot\mid i\right)\bmod n$ are zero, leaving only $\nicefrac{p s(s+1)}{2}$ of the original $\nicefrac{n s (s+1)}{2}$ equations.

\section{Conclusion}\label{sec:conc}

After a short recapitulation of the well-known re-encoding map, we have introduced the periodicity projection as a special case and have given its most important properties. An analysis of the interpolation step of the GSA has shown that applying either of the maps to the received vector results in a significant compression of the involved homogeneous system of linear equations. We have shown that compression and decompression are particularly simple in case of the periodicity projection, since the involved polynomials have large palindromic factors that can be calculated in advance. Besides further elaboration of the computational savings, it appears interesting to investigate the factorization step of the GSA when provided with the highly structured bivariate result polynomial in case of the periodicity projection.



\vspace{0.6cm}


\end{document}